\documentstyle[12pt,aasms4,psfig]{article}
\newcommand{\ASCA}{{\it ASCA}}
\newcommand{\Granat}{{\it Granat}}
\newcommand{\Einstein}{{\it Einstein}}
\newcommand{\ROSAT}{{\it ROSAT}}
\newcommand{\etal}{et al.}
\newcommand{\NH}{{$N_{\rm H}$}}
\newcommand{\NHUNIT}{${\rm H}\ {\rm cm}^{-2}$}
\newcommand{\NFe}{{$N_{\rm Fe}$}}
\newcommand{\NFeUNIT}{${\rm Fe}\ {\rm cm}^{-2}$}
\newcommand{\FX}{{$F_{\rm X}$}}
\newcommand{\FLUXUNIT}{${\rm erg}\ {\rm cm}^{-2}\ {\rm s}^{-1}$}

\newcommand{\LUMIUNIT}{${\rm erg}\ {\rm s}^{-1}$}
\newcommand{\GA}{1E~1740.7$-$2942}

\newcommand{\hms}[3]{$#1^{\rm h}#2^{\rm m}#3^{\rm s}$}

\newcommand{\dmnos}[2]{$#1^{\circ}#2'$}

\lefthead{Sakano et al.}
\righthead{Further studies of {\GA} with {\ASCA}}

\begin{document}

\title{Further studies of {\GA} with {\ASCA}}

\author{Masaaki Sakano\altaffilmark{1}, Kensuke Imanishi,  Masahiro Tsujimoto, Katsuji Koyama\altaffilmark{2}}
\affil{Department of Physics, Kyoto University,
 Kyoto 606-8502 Japan,
 sakano@cr.scphys.kyoto-u.ac.jp, kensuke@cr.scphys.kyoto-u.ac.jp,
 tsujimot@cr.scphys.kyoto-u.ac.jp, koyama@cr.scphys.kyoto-u.ac.jp}

\and

\author{Yoshitomo Maeda\altaffilmark{1}}
\affil{Department of Astronomy and Astrophysics,
 Pennsylvania State University,
 University Park PA 16802-6305 U.S.A.,
 maeda@astro.psu.edu}

\altaffiltext{1}{Research Fellow of the Japan Society for the Promotion 
of Science.}
\altaffiltext{2}{CREST: Japan Science and Technology Corporation (JST)}

\begin{abstract}

    We report the {\ASCA} results of the Great Annihilator {\GA} obtained
 with five pointing observations in a time span of 3.5 years.
   The X-ray spectrum for each period is
 well fitted with a single power-law absorbed by a high column of gas.
   The X-ray flux changes by a factor of 2 from period to period,
 but the other spectral parameters show no significant change.
   The photon index is flat with $\Gamma=$ 0.9--1.3.
   The column densities of hydrogen {\NH} is
 $\sim 1.0\times 10^{23}$ {\NHUNIT} and
 that of iron {\NFe} is $\sim 10^{19}$ {\NFeUNIT}.
   These large column densities indicate that {\GA} is near
 at the Galactic Center.  The column density ratio leads
 the iron abundance to be 2 times larger than the other elements
 in a unit of the solar ratio.
   The equivalent width of the K$\alpha$-line from a neutral iron
 is less than 15 eV in 90\% confidence.
   This indicates that the iron column density within several parsecs
 from {\GA} is less than $5 \times 10^{17}$ {\NFeUNIT}.
   In addition, the derived hydrogen column density is about 1/6 of
 that of giant molecular clouds in the line of sight.
   All these facts support that {\GA} is not in a molecular cloud,
 but possibly in front of it;
 the X-rays are not powered by accretion from a molecular cloud,
 but from a companion star like ordinary X-ray binaries.

\end{abstract}


\section{Introduction \label{sec:1740-intro}}

	The center of our Milky Way Galaxy is one of the most complex regions
 in X-rays as well as the other wave bands.  In fact, many X-ray point sources,
 including transients, have been detected with the past observations.
   Among these, {\GA} has been found to be quite unusual.   This source was
 discovered with the {\Einstein}/IPC in the soft X-ray band below 3.5 keV
 (\cite{Hertz84}).  Higher energy observations revealed an unusually hard
 spectrum (Skinner {\etal} 1987, 1991; \cite{Kawai88}; \cite{Cook91};
 \cite{Bazzano92}; Cordier {\etal} 1993a, 1993b; \cite{Goldwurm94}).
   In fact, this is by far the brightest source within a few degree of
 the Galactic Center in the high energy band
 above 20 keV with a power-law spectrum extending up to 100 keV or higher.
    It showed flux variability in the time scale from a day to a few years 
 (\cite{Bouchet91}; \cite{Sunyaev91}; Churazov {\etal} 1993a, 1993b;
 \cite{Paciesas93}; Pavlinsky, Grebenev, \& Sunyaev 1994;
 Zhang, Harmon, \& Liang 1997).
   Variability of time scale shorter than a day has been detected
 by Smith {\etal} (1997).
   The spectral shape has been almost stable in time, regardless of
 the flux variations.
   These spectra and time variabilities generally
 resemble those of Cygnus  X-1 (e.g., \cite{Kuznetsov97}),
 the prototype of the galactic black hole candidate.
	Unlike Cygnus X-1, however, optical and IR observations have failed
 to find any companion stars
 (\cite{Prince91}; \cite{Mereghetti92}; \cite{Djorgovski92}),
 probably due to strong extinction to the direction of the Galactic Center.

   	The electron-positron annihilation line at 511-keV from this source
 was reported with the {\Granat}/SIGMA observations, hence {\GA} has been
 referred to as the ``Great Annihilator'' (\cite{Bouchet91}; \cite{Sunyaev91};
 \cite{Churazov93b}; \cite{Cordier93a}).
   Jung {\etal} (1995), Harris, Share, \& Leising (1994a, 1994b), Malet {\etal} (1995),
 Smith {\etal} (1996), Cheng {\etal} (1998), on the other hand, reported no evidence for the 
 annihilation line, making this issue debatable.
   The VLA radio observations revealed that {\GA} has a radio counterpart
 with variable flux correlated to the X-ray variation
 (Mirabel {\etal} 1992, 1993).  They also found non-thermal 
 double jet-like structures emanating from the source,
 which resemble those of the galactic superluminal jet sources
 (e.g., Fender, Burnell, \& Waltman 1997).
   Hence this source is also referred to as a ``micro quasar''. 
  These facts altogether place {\GA} to be one of the best candidates
 of stellar mass black holes.

  The CO observations found a giant molecular cloud whose density peak
 is in agreement with the position of {\GA} (\cite{Bally91};
 \cite{Mirabel91}).
  They proposed that {\GA} is an isolated or a binary black hole,
 powered directly from a surrounding molecular cloud
 such as by the Bondi-Hoyle accretion (\cite{Bondi44}),
 although the conventional binary
 scenario with mass accretion from a companion star cannot be excluded.

	X-ray spectrum and time variability would be
 keys for the study of the nature and emission mechanisms.
  In particular, the equivalent width of fluorescent iron line as a function of
 absorption column provides direct evidence whether the source is really
 in a dense cloud or not.
  Churazov, Gilfanov, \& Sunyaev (1996) and Sheth {\etal} (1996) have examined
 the X-ray spectra of this source with the Performance Verification (PV)
 and AO-2 phases of the early {\ASCA} operations.
   Their derived {\NH} values, hence the conclusions,
 are not fully consistent with each other.
  This is partly due to the fact
 that they fitted the spectra with different energy range.
  More serious problem
 for the detailed spectral study of this source, particularly
 on the iron line, iron K-edge and low energy absorption structures,
 is possible contamination of the thin thermal Galactic plasma
 emission which includes strong emission lines (\cite{Koyama96}).
  Thus we have made further observations in AO-3 and AO-5 phases
 after the PV and AO-2,
 analyzed all the available data sets, putting
 our particular effort to remove properly the possible contamination
 from the plasma emission in the Galactic Center region.
    We assume the distance of the Galactic Center to be 8.5 kpc.

\section{Observations \label{sec:1740-obs}}

	{\ASCA} observations of {\GA} were made on the five occasions,
 which are listed in Table~\ref{tbl:obslog}.
   Two sets of the Solid-State Imaging Spectrometers (SISs)
 were operated in parallel with
 the Gas Imaging Spectrometers (GISs),
 both located separately at foci of four identical thin-foil X-ray telescopes (XRTs).   
  Details of this instrumentation are
 found in Serlemitsos {\etal} (1995), Burke {\etal} (1991, 1994), Yamashita {\etal} (1997),
 Ohashi {\etal} (1996), Makishima {\etal} (1996), while a general description
 of {\ASCA} can be found in Tanaka, Inoue, \& Holt (1994).

   In the PV and AO-2 observations, {\GA} was pointed near the center
 of the field of view,
 while for the other observations, {\GA} was out of the SIS field of view.
   In addition,  we have made 8 sequential pointing observations
 near the Galactic Center with 20 ksec exposure for each.  
These series of observations covered the Galactic Center region of
 about $1\times 1.5$ degree$^2$ and the data were used to estimate
 the Galactic diffuse background (see Sec.~\ref{sec:1740-variability}).
	Table~\ref{tbl:obslog} summarizes
 the observation log. 
Data reduction and cleaning were made with the standard method
 as described in the user guide by NASA Goddard Space Flight Center (GSFC).
The event selection and the analysis were performed
 on UNIX workstations with the FTOOLS and XANADU packages released
 from the NASA GSFC.
 
\placetable{tbl:obslog}

\section{Results and Analysis \label{sec:1740-result}}

\subsection{X-ray images \label{sec:1740-image}}

\begin{figure}
\centering
 \centerline{\psfig{file=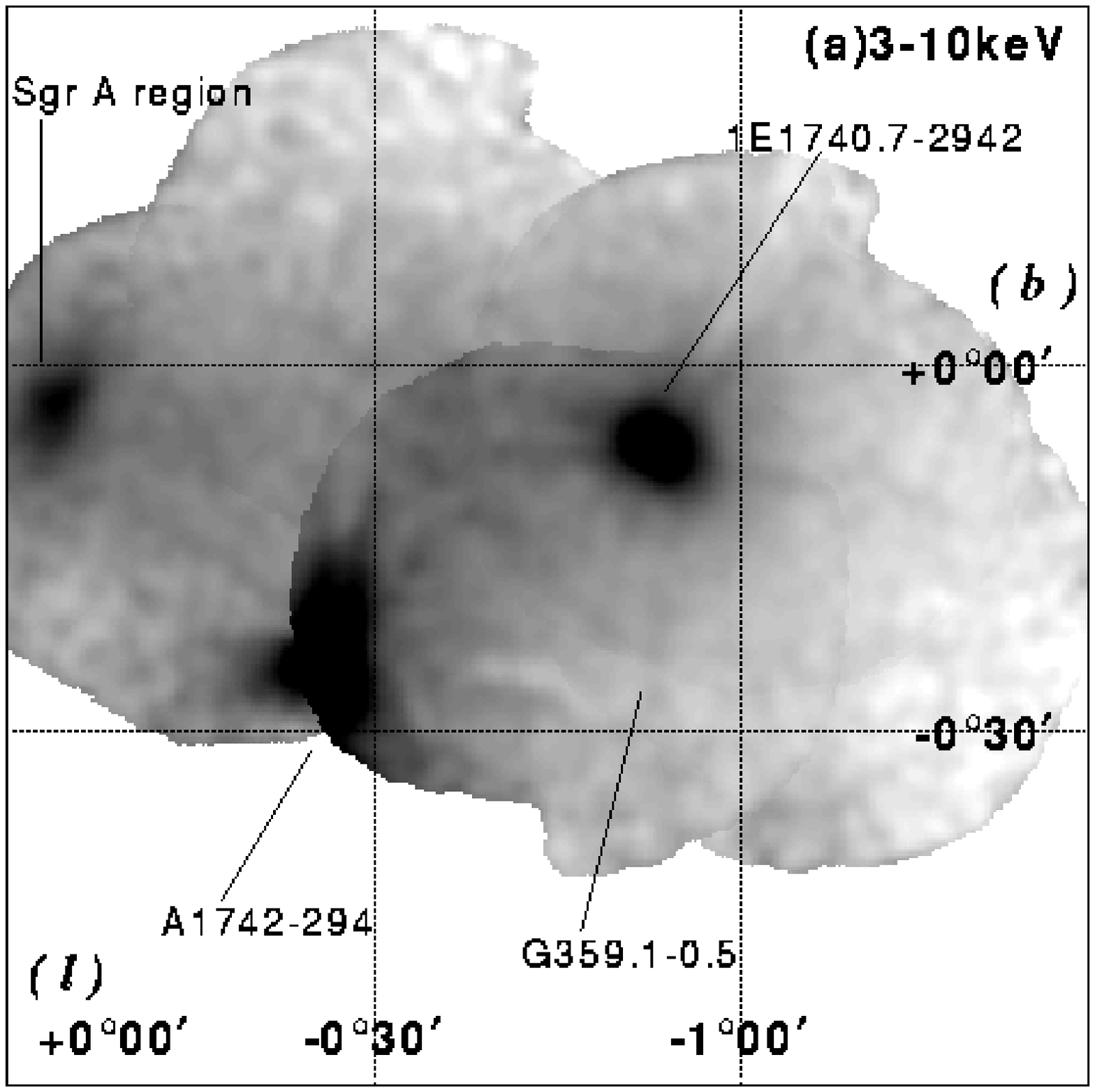,width=\textwidth,clip=}}
  \caption[f1a.eps,f1b.eps]
{The  GIS mosaic images near {\GA}; (a) in the 3--10 keV band,
 (b) in the 0.7--1.5 keV band.
 The images are taken with multiple pointings of {\ASCA}
 in 1993 -- 1997.  The data of GIS2 and GIS3 are summed,
 smoothed with a Gaussian filter of $\sigma =$ 3 pixels
 ($\sim$ 0.75 arcmin),
 and corrected for exposure, vignetting and the detection efficiency
 with GIS grid,
 after non X-ray background is subtracted.
  The dotted grids are the galactic coordinate ($l_{\rm II}$, $b_{\rm II}$).
 Color levels are logarithmically spaced.
   The stray lights by some bright sources, which are variable from
 epoch to epoch, make images complicated,
 for example, they make images discontinuous on the edges of the fields
 of views.   A faint soft source,
 AX J1744.3$-$2940, at ($17^{\rm h}44^{\rm m}18^{\rm s}$, $-29^{\circ}40.\!'1$)
 in J2000 coordinate is found.  The positions of Sgr~A region and
 SNR G359.1$-$0.5 (\cite{Yokogawa98}) are also indicated.

\label{fig:gc-image}
}
\end{figure}

  	The GIS mosaic images for the hard X-ray band (3--10 keV)
 and the soft X-ray band (0.7--1.5 keV)
 are given in Figure~\ref{fig:gc-image}a and \ref{fig:gc-image}b,
 respectively.
   We find a strong point source at the {\ROSAT} position of {\GA} 
 ((17$^{\rm h}$43$^{\rm m}$54$^{\rm s}\!$.8,  $-$29$^{\circ}$44$'$38$''$)
 in J2000 coordinate (Heindl, Prince, \& Grunsfeld 1994)),
 within the nominal {\ASCA} error circle of about 1$'$ radius in each image.
  The excess X-ray flux above the diffuse background level is higher
 in the hard X-ray band than that in the soft band.
  This immediately indicates that either the X-rays are highly absorbed,
 or the spectrum is very hard, or probably both.

   We also find a faint soft source, AX J1744.3$-$2940,
 at ($17^{\rm h}44^{\rm m}18^{\rm s}$, $-29^{\circ}40.\!'1$)
 in J2000 coordinate with the error circle of about 1$'$.
  This position is in good coincidence with the source~2 in Heindl {\etal}
 (1994) observed with {\ROSAT}, or 1WGA J1744.2$-$2939.
  It is about 7$'$ offset from {\GA},
 hence the photon contamination from this source for the analysis
 of {\GA} is negligible.

\placefigure{fig:gc-image}

\subsection{Light curves \label{sec:lightcurve}}

   Since the X-rays from {\GA} are limited to above about 2 keV
 (see next section or X-ray images in Figure~\ref{fig:gc-image}),
 we made light curve for each observation, accumulating 
 2--10 keV photons in the circular region
 within 3$'$ radius from the source.
  To increase statistics, we summed the data of SIS 0 and 1,
 or GIS 2 and 3 with a time bin of 256 sec.  
  As an example, we show the light curves of the AO-2 1st observation
 in Figure~\ref{fig:lightcurve}.  
  Constant flux assumption for the light curve of each observation
 is rejected by $\chi^2$-tests with more than  90\% confidence.
  Thus  the X-ray flux was found to be variable 
although the variability amplitude is not large.

  We also examined FFT analysis for the GIS 2$+$3 data of high-bit rate
 with 1/16 sec time resolution, and found no periodic variation
 in the time scale of 1/16 -- 1000 second
 from any of the five separate observations.  

  As for the long term variability, we found the averaged flux
 in PV phase decreased to about 75\% and 50\% in AO-2 and AO-3 phases,
 respectively, and then increased in AO-5 phase to about the same flux
 as in PV phase (see Figure~4 and Sec.~3.3).

\placefigure{fig:lightcurve}

\subsection{Spectral Analysis  \label{sec:1740-variability}}
 
    	{\GA} is located near the Galactic Center region, which is filled with a high temperature plasma.  
The plasma emits  diffuse X-rays including strong iron lines, 
 with significant flux variations from position to position (\cite{Koyama96}).
Therefore  the spectrum of {\GA} may  be   contaminated by  the diffuse X-rays,  especially in the low energy 
band and in the  iron line feature. 
   
	To minimize possible effect to the spectrum due to
 the background variation, we accumulated the source X-ray photons in a small
 circle of 2$'$ radius and extracted the background taken from the annulus
 of 2$'$--4$'$ radius.   We thus made the background subtracted spectra
 with GIS2$+$3 and SIS0$+$1 for each observation phase.
   For the AO-3 data, we excluded the region closely near to the detector edge
 or the calibration isotope from the background region.
	
    We fitted each spectrum with an absorbed power-law, using the photo-electronic
 cross-sections by Ba{\l}uci\'{n}ska-Church \& McCammon (1992).  The abundances
 for the absorbing matters were fixed to the solar values (\cite{Anders89}),
 but that of iron was allowed to be varied, because the 7.1-keV edge structure
 carries unique information of the iron column.
  Examples (AO-2) of the spectra with the best-fit models are
 given in Figure~\ref{fig:fit-vabs} and the best-fit parameters
 for all the observations are listed in Table~\ref{tbl:fit-vabs}.

 We then fitted the GIS and the SIS data for each phase simultaneously
 with a power-law model.   The results with the best statistics are
 also given in Table~\ref{tbl:fit-vabs}.
  The best-fit photon index is in the range of 0.9--1.3,
 the hydrogen column density is (8--11)$\times 10^{22}$ {\NHUNIT},
 and the iron abundance is found to be about 2 solar.

    Although we found no clear emission line in each spectrum,
 we included a 6.4~keV line for the further fitting,
 because the emission line from neutral iron at 6.4 keV gives
 direct  information around the cold gas near {\GA}.
  In this fitting we fixed the line 
 energy and the intrinsic width to be 6.4 keV and 0 keV, respectively.
  The resultant parameters are listed in Table~\ref{tbl:fit-vabs}
 together with the 2--10 keV flux (absorption included).

\placetable{tbl:fit-vabs}
\placefigure{fig:fit-vabs}

    In Figure~\ref{fig:long-term}, we summarize the long term
 spectral behavior.  No significant difference, except for
 the flux change, is found from observation to observation
 in a time span of 3.5 years.  In addition, the spectral slope is rather flat.
  Thus we can conclude that {\GA} has been in the hard sate
 of the galactic black hole spectrum.

\placefigure{fig:long-term}

  Since the galactic black hole candidates often have a soft component
 with temperature of less than 1.0 keV 
 (e.g., \cite{Tanaka95}; \cite{Paradijs98}),
 we examined whether the {\GA} spectrum requires a soft component or not.
We tried to fit the data with a power-law model adding a soft black body
 component, and found no significant improvement by this procedure,
 hence the spectrum requires no soft component.
However it is not clear whether this is real or merely due to
 large absorption in the low energy band.

The Galactic diffuse emission includes strong iron K-shell lines
 from neutral (at 6.4~keV) and highly ionized irons (at 6.7~keV and 6.97~keV)
 (\cite{Koyama96}; Maeda 1998).
  Since the K-shell line of a neutral iron (6.4~keV) plays a key role for
 the nature of {\GA}, 
 we estimated the fluctuation of the diffuse K-shell lines
 (the 6.4~keV-lines)
 in the source extraction region, using the SIS data.
   We made spectra from a circular region with a radius of 4$'$
 near the center of each SIS chip in the mapping observations
 of the Galactic Center region.
  We then fitted each spectrum in the 4.5--10 keV band
 with a continuum plus three Gaussian lines fixing  the center energies 
 to be 6.4, 6.7 and 6.97 keV.
  The intensity ratio of the 6.7~keV to 6.97~keV lines
 was kept to be constant from place to place, because Maeda (1998) and
 Koyama {\etal} (1996) found no significant variation of the flux ratio.
    Figure~\ref{fig:iron-diffuse} shows the 6.4~keV-line fluxes along  positions, where the flux  
 is converted to the value in a 2$'$ radius, the same radius of the source region.
   From Figure~\ref{fig:iron-diffuse}, we can safely conclude that
 uncertainty of the iron line flux in {\GA}
 due to the background fluctuation
 is less than $0.6\times 10^{-5}$ photon~sec$^{-1}$~cm$^{-2}$/(2$'$-circle).
  This value is, at most, 15\% of that from {\GA}, hence can be ignored
 in the following discussion. 

\placefigure{fig:iron-diffuse}

\section{Discussion \label{sec:1740-discuss}}

\subsection{Absorption and Iron Abundance}

	The {\ASCA} results of {\GA} have been already reported
 by Churazov {\etal} (1996) and Sheth {\etal} (1996).
  Although they used essentially the same data sets, their results are
 not fully consistent with each other.
  The major difference is found in the {\NH} values. 

	With a single power-law fitting and absorption gas
 of solar abundance, Churazov {\etal} (1996) gave
 {\NH} to be $1.7\times 10^{23}$ {\NHUNIT} 
 of the {\ASCA} spectrum in the range of the 4--10 keV band,
 but found $1 \times 10^{23}$ {\NHUNIT} when the full energy range of 0.4--10 keV
 was used. 
   Sheth {\etal} (1996) estimated {\NH} to be $0.8\times 10^{23}$ {\NHUNIT}
 by the same model fitting but in the 0.5--12 keV range.  We found that
 the apparent inconsistency can be solved if iron is overabundant
 relative to the other elements.  In fact, we have shown that
 the spectra of {\GA} are well presented with a single-power model
 absorbed by the gas column of $1 \times 10^{23}$ {\NHUNIT},
 in which iron abundance is 2 solar.

    Murakami {\etal} (1998) analyzed the reflected X-ray
 by the giant molecular cloud Sgr~B2, which is located in nearly
 the same (angular) distance but in the opposite direction from
 the Galactic Center, and found that iron is overabundant
 relative to the other elements.
  On the other hand, Sellgren, Carr, \& Balachandran (1997) and Ramirez {\etal} (1998),
 with the infrared observations of the atmosphere
 on the giant stars near the Galactic Center,
 estimated that the iron abundance relative to hydrogen is consistent
 with the solar value. 
   Thus further deep observations of K-edge absorption feature
 of X-ray sources near the Galactic Center are required.

	Sakano {\etal} (1999) analyzed many X-ray sources
 located near the Galactic Center region, and found that
 the {\NH} values can be well described by a simple function
 of the galactic latitude (Figure~2 in Sakano {\etal} (1999)).
  From their results, we estimate that the interstellar absorption
 to the {\GA} direction is  (1--2)$\times 10^{23}$ {\NHUNIT},
 in agreement with the best-fit {\NH} of the {\GA} spectrum.
	No change of {\NH} during the {\ASCA} long term observations
 implies that the absorption region should have a size larger than
 a few light-yrs, excluding a possibility of an accretion disk
 or stellar corona.
   Thus {\GA} is very likely to be located near at the Galactic Center.

  	The radio observations of the large molecular cloud toward
 {\GA}, on the other hand, gave
 the total hydrogen column density to be {\NH}$\sim 6\times 10^{23}$ {\NHUNIT}
 (\cite{Bally91}; \cite{Vilhu96}).   One may argue, however,
 that the conversion factor
 from the CO line intensity to the hydrogen column density may have
 large uncertainty.   Oka {\etal} (1998), for example, pointed out
 that the usual conversion factor is not appropriate for the clouds
 in the Galactic Center region, and that the derived mass
 in the ordinary manner should be reduced by several factors.  
   Taking this possible uncertainty into account,
 we still suspect that the molecular cloud is located behind {\GA}.   

\subsection{6.4-keV iron line \label{sec:diffuse-6.4}}

    We found no significant line emission from a neutral iron (6.4~keV-line).
  The upper-limits of the equivalent width are estimated to be a few ten eV,
 depending on the observation period and detector.
   The most stringent limit is found to be 15 eV
 in the combined (GIS and SIS) analysis of AO-2.
   These results of small equivalent width further confirm
 the result by Churazov {\etal} (1996) with larger data sets.  

   The equivalent width $<$15~eV implies that
 the mean iron column density around {\GA} is less than
 {\NFe}$< 5 \times 10^{17}$ {\NFeUNIT} (\cite{Inoue85}; \cite{Awaki91}).
    This value is smaller than 1/10 of the total column density estimated
 from the 7.1 keV edge depth.
  Hence {\GA} cannot be in a local dense cloud as suggested
 by Bally \& Leventhal (1991) and Mirabel {\etal} (1991).   

    Bally \& Leventhal (1991) estimated the required local density
 of the interstellar matter around {\GA} for the observed luminosity
 of {\GA} when it is powered directly
 by the interstellar matter in the case of Bondi-Hoyle accretion
 (\cite{Bondi44}).  The calculated luminosity is,
\begin{equation}
  L = 2.3\times 10^{36}~\eta \left[\frac{M}{M_{\odot}}\right]^{2}
	\left[\frac{v}{10~{\rm km~s^{-1}}}\right]^{-3}
	\left[\frac{n_{\rm H}}{10^4~{\rm cm^{-3}}}\right]\ \ ({\rm erg~s^{-1}}),
  \label{eq:bondi}
\end{equation}
 where $\eta$ is the fraction of the released energy with emission
 from the rest mass energy,
 $M$ is the mass of the central compact object,
 $v$ is the relative velocity of the compact object against the cloud, and
 $n_{\rm H}$ is the hydrogen density of the cloud.
   The highest luminosity of {\GA} in the {\ASCA} observations is
 estimated to be $3 \times 10^{36}$ {\LUMIUNIT},
 from the X-ray flux of $2 \times 10^{-10}$ {\FLUXUNIT}
 in the 2--10 keV band and the source distance of 8.5 kpc.
     Since the spectrum is extracted within 2$'$-radius (5~pc-radius),
 the observed upper-limit of the iron equivalent width
 is converted to the local density of
\begin{equation}
 n_{\rm H}= 7\times 10^{2}~(Z_{\rm Fe})^{-1}\ \ ({\rm H}~{\rm cm}^{-3}),
\end{equation}
 where $Z_{\rm Fe}$ is iron abundance relative to solar.

   Then, assuming $\eta=$ 0.1, we can set possible ranges
 of $M$ and $v$ for the Bondi-Hoyle accretion scenario
 as is given in Figure~\ref{fig:bondi-prm} with solid lines, while the 
 dashed line shows the most probable case of $Z_{\rm Fe}=2$ in equation (2).
   Thus we find that the velocity of {\GA} must be less than 10 km~s$^{-1}$
 even in the case of the large mass of 20${\rm M}_{\odot}$
 under the Bondi-Hoyle accretion scenario.

    The velocity of black holes would be 
 larger than that of normal stars due to additional kick velocity  
 by supernova explosions. 
  Since no data of the velocity dispersion of black holes are found,
 the best alternative way to estimate the velocity of {\GA} 
 is to use the velocity dispersion of neutron stars, 
 which are also produced by supernova explosions. 
  Thus we refer the velocity dispersion of the radio pulsars (neutron stars)
 (\cite{Lorimer97}; \cite{Hansen97}) and find that the velocity of {\GA} would
 be larger than 10 km~s$^{-1}$ with more than 99\% probability.
  We therefore conclude that the Bondi-Hoyle scenario that {\GA} is 
 powered directly from a molecular cloud is unlikely.

\placefigure{fig:bondi-prm}

\subsection{Long term variability \label{sec:1740-d-variability}}

 	Pavlinsky {\etal} (1994) reported the long term variability of {\GA}
 measured with ART-P on board {\Granat}.  According to their results,
 {\GA} was usually in a medium or high state since 1990 with occasional
 exceptions of low state.   
  Since the spectral shape during the {\ASCA} observations
 showed no significant change except for the variability
 of the total flux, we safely assume that the spectra in the {\Granat}
 observations has the same photon index ($\Gamma\sim$ 1.2)
 as those of the {\ASCA} spectra.
   Then we found that the {\ASCA} fluxes of
 (1--2)$\times 10^{-10}$ {\FLUXUNIT} in the 2--10 keV band
 corresponds to (2.5--5)$\times 10^{-10}$ {\FLUXUNIT} in the 8--20 keV band,
 the middle to high-state fluxes of {\Granat}.
   The {\ASCA} flux is also in agreement with the BATSE light curve
 for the same period (\cite{Zhang97}).
   The similarity between the {\ASCA} and BASTE light curves
 is another indication of the spectral invariance over larger energy range.

\section{Summary \label{sec:1740-summary}}

        Using all the {\ASCA} data of {\GA}, available at the time of
 this writing, we conclude as follows:

\begin{enumerate}
\item   We found that {\GA} was in the middle to high flux state, showing
 long term variation of the factor of two in a span of 3.5 years, 
 while the photon index and column density have been nearly constant.

\item   The wide band spectrum in the 1--10 keV band is well fitted
 with an absorbed power-law,  where the iron abundance is
 twice as the others in a unit of solar value.
  The power-law slope is particularly flat with $\Gamma =$ 0.9--1.3,
 which implies that {\GA} was in the hard state.
  The large hydrogen column density of {\NH}$\sim 1\times 10^{23}$ {\NHUNIT}
 supports the location of {\GA} to be near the Galactic Center.

\item   From the iron-edge structure we estimated the iron column density
 to be {\NFe} $\sim 1\times 10^{19}$ {\NFeUNIT}.

\item  The iron equivalent width has been very small with the upper-limit
 of 15~eV.   This indicates that the iron column near around {\GA} is
 less than {\NFe} $< 5 \times 10^{17}$ {\NFeUNIT}.

\item  Our results favor the geometry that {\GA} is not in the radio molecular
 cloud. The cloud would be, by chance, in the line of sight to, or
 possibly behind {\GA}.  This indicates that the X-ray emission
 does not originate by the direct accretion from the 
 molecular cloud.  {\GA} is probably a normal black hole binary.

\end{enumerate}

\acknowledgments

    The authors express their thanks to all of the members of the {\ASCA} team
 whose efforts made these observations and data analysis possible.
  We are grateful to Dr. S.~Yamauchi, Prof. Ph.~Durouchoux, and,
 particularly, an anonymous referee for their valuable comments
 and suggestions.
  M. S. thanks to Drs. H.~Sogawa, M.~Ozaki, Y.~Fujita and H.~Matsumoto
 for discussion.
  M. S. and Y. M. acknowledge the supports from
 the Japan Society for the Promotion of Science for Young Scientists.

\clearpage

\addtocounter{figure}{-1}

\begin{figure}
\centering
\vspace*{1cm}

 \centerline{\psfig{file=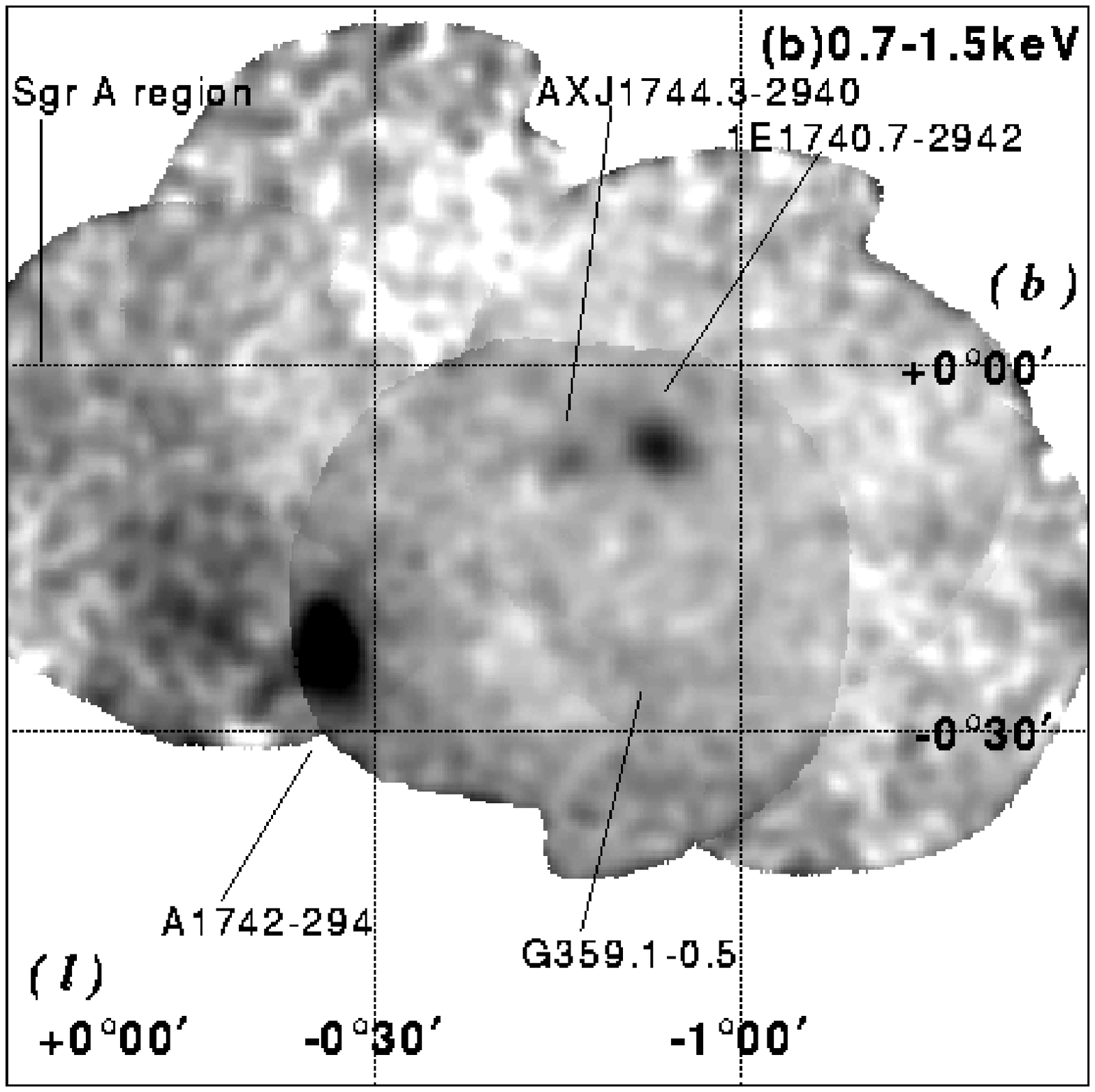,width=\textwidth,clip=}}
   \caption[\ \ \ (continued)]{(b)}
\end{figure}

\begin{figure}
\centering
\centerline{\psfig{file=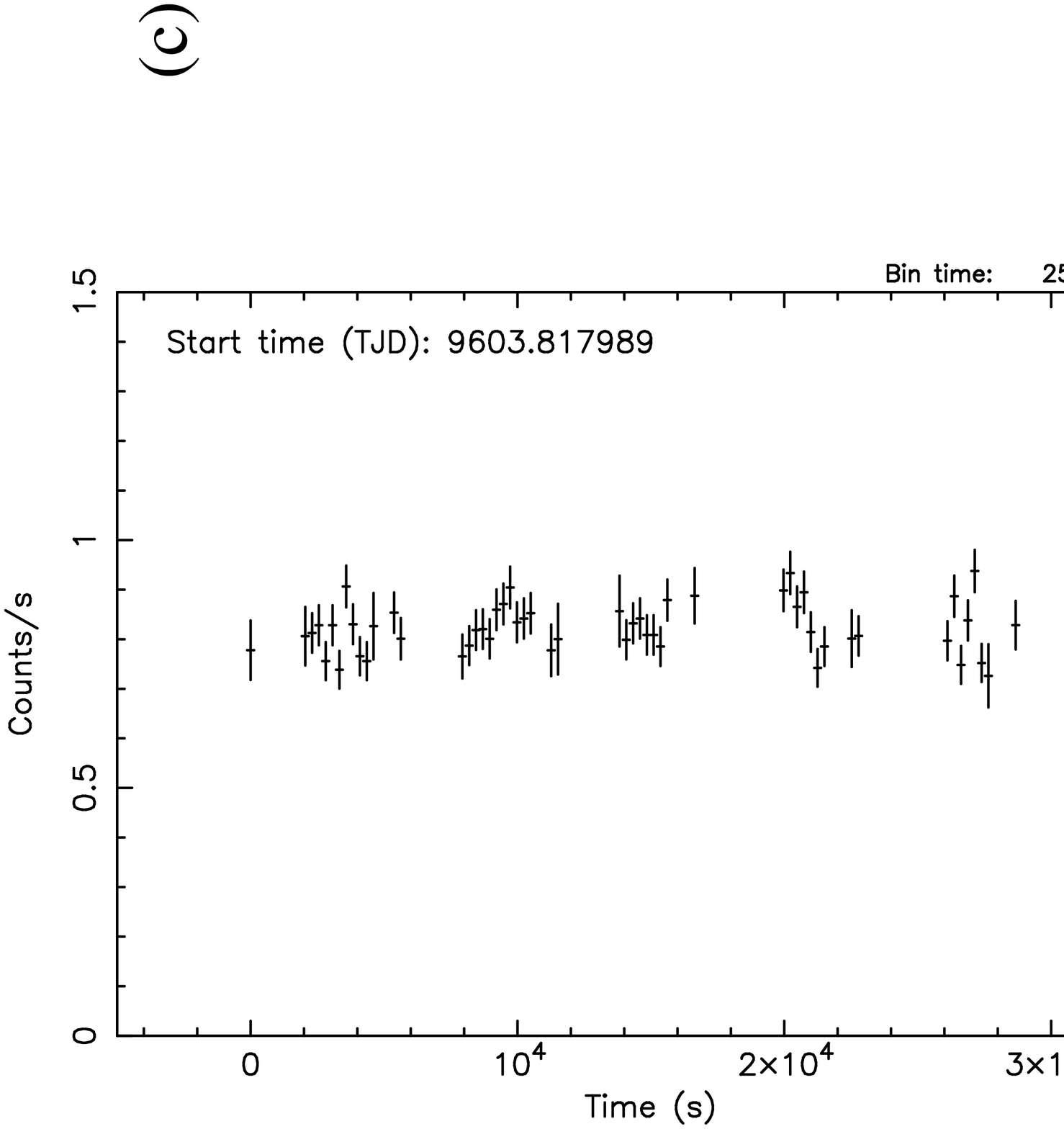,width=0.85\textwidth,clip=}}
\centerline{\psfig{file=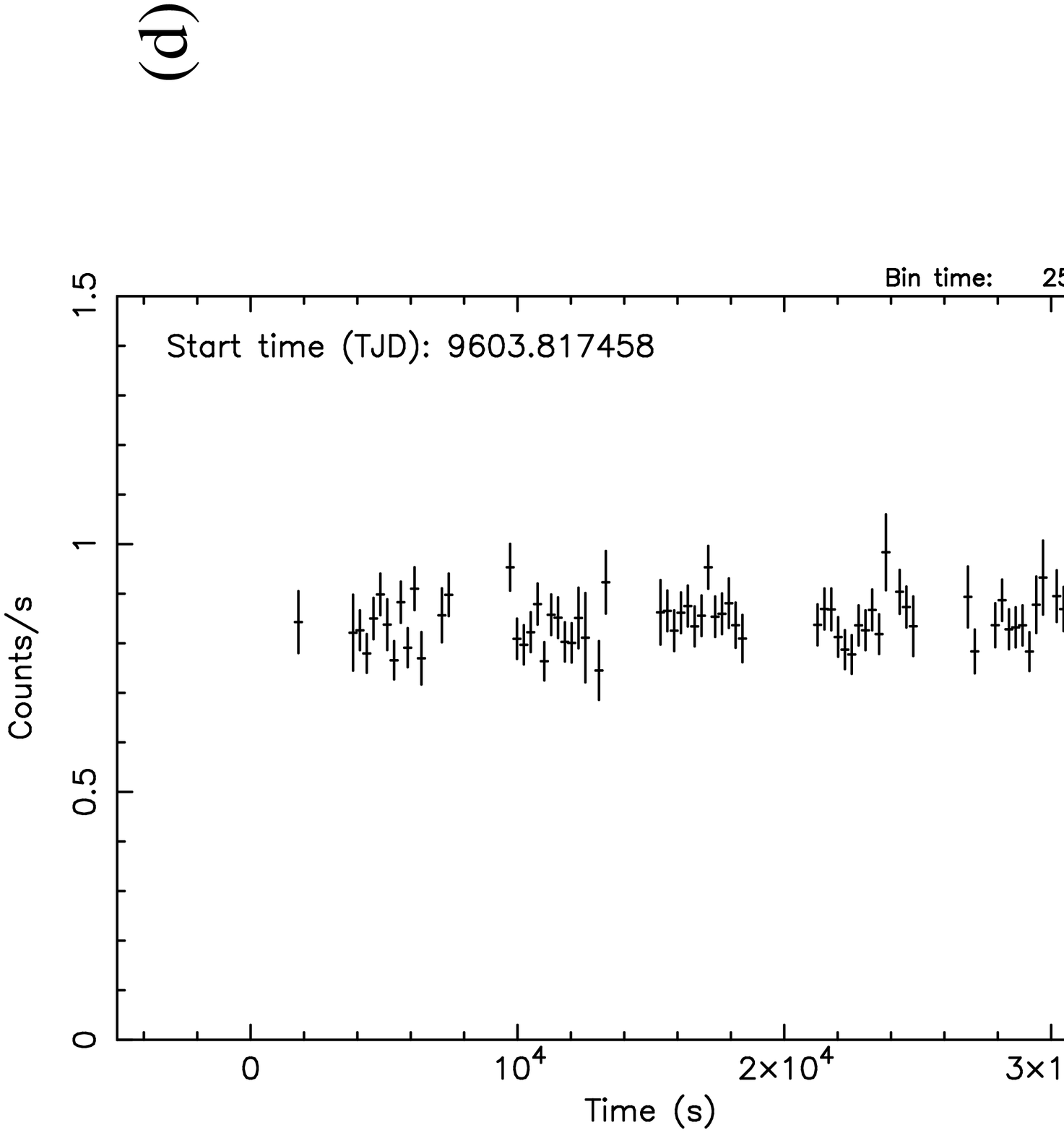,width=0.85\textwidth,clip=}}
  \caption[f2a.eps,f2b.eps]
{The 2--10 keV light curves with 256~second bin for the AO-2 1st observation:
 (left) GIS~2$+$3 data and  (right) SIS~0$+$1.
\label{fig:lightcurve}
}

\end{figure}

\begin{figure}
\centering
\centerline{\psfig{file=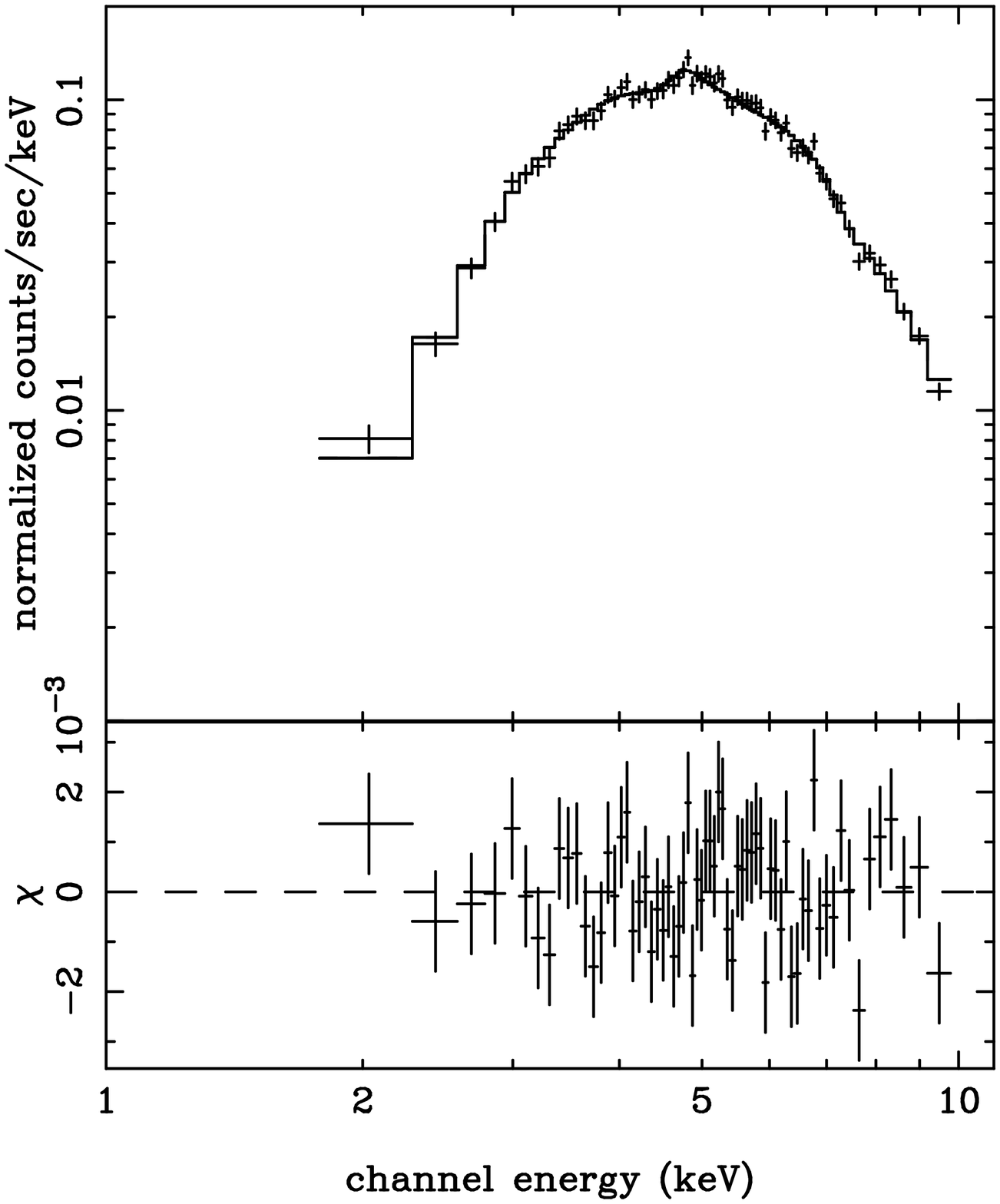,width=0.48\textwidth,clip=}}
\centerline{\psfig{file=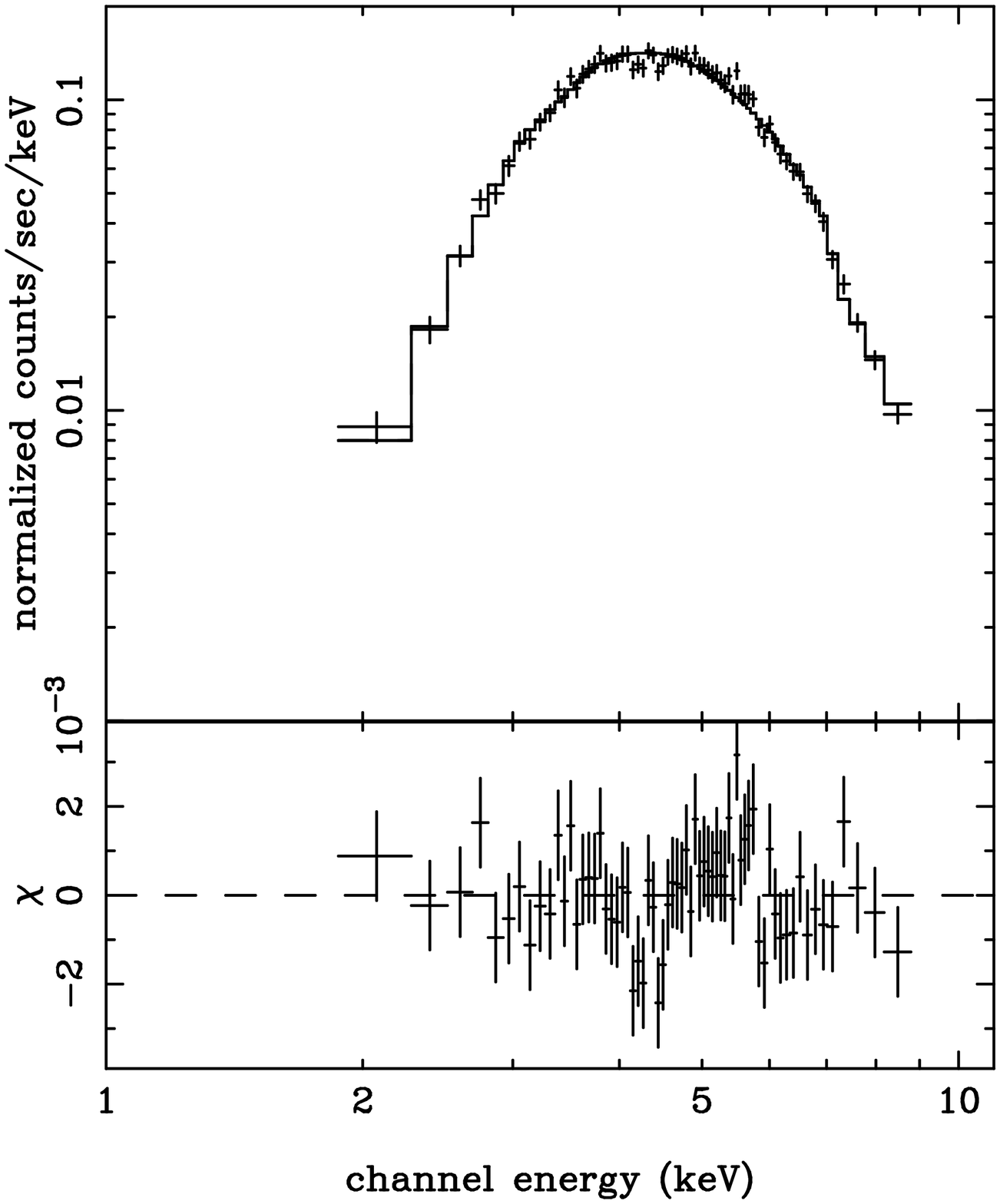,width=0.48\textwidth,clip=}}
  \caption[f3a.eps,f3b.eps]
{The spectra of AO-2 with the best-fit model (solid lines): 
(left) GIS~2$+$3  and  (right) SIS~0$+$1.
\label{fig:fit-vabs}
}
\end{figure}

\begin{figure}
\centering
  \centerline{\psfig{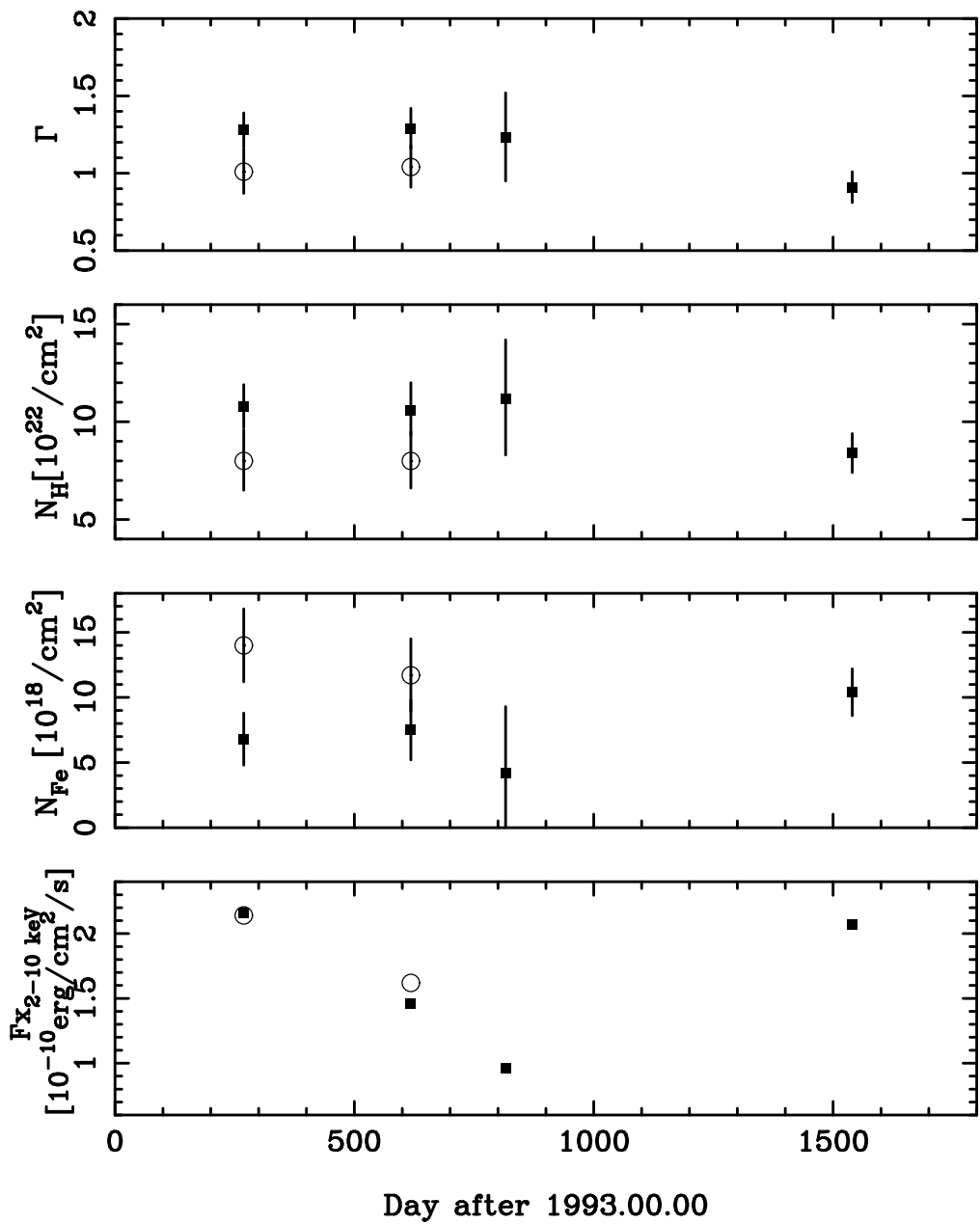}}
  \caption[f4.eps]
{Long term variability of a power-law photon index $\Gamma$,
 hydrogen column density {\NH} in unit of $10^{22}$ {\NHUNIT},
 iron column density {\NFe} in unit of $10^{18}$ {\NFeUNIT} and
 the X-ray flux in the 2--10 keV band {\FX} in unit of $10^{-10}$ {\FLUXUNIT}.
Close boxes and open circles represent the data taken in GIS and SIS, respectively.

\label{fig:long-term}
}
\end{figure}

\begin{figure}
\centering
  \centerline{\psfig{file=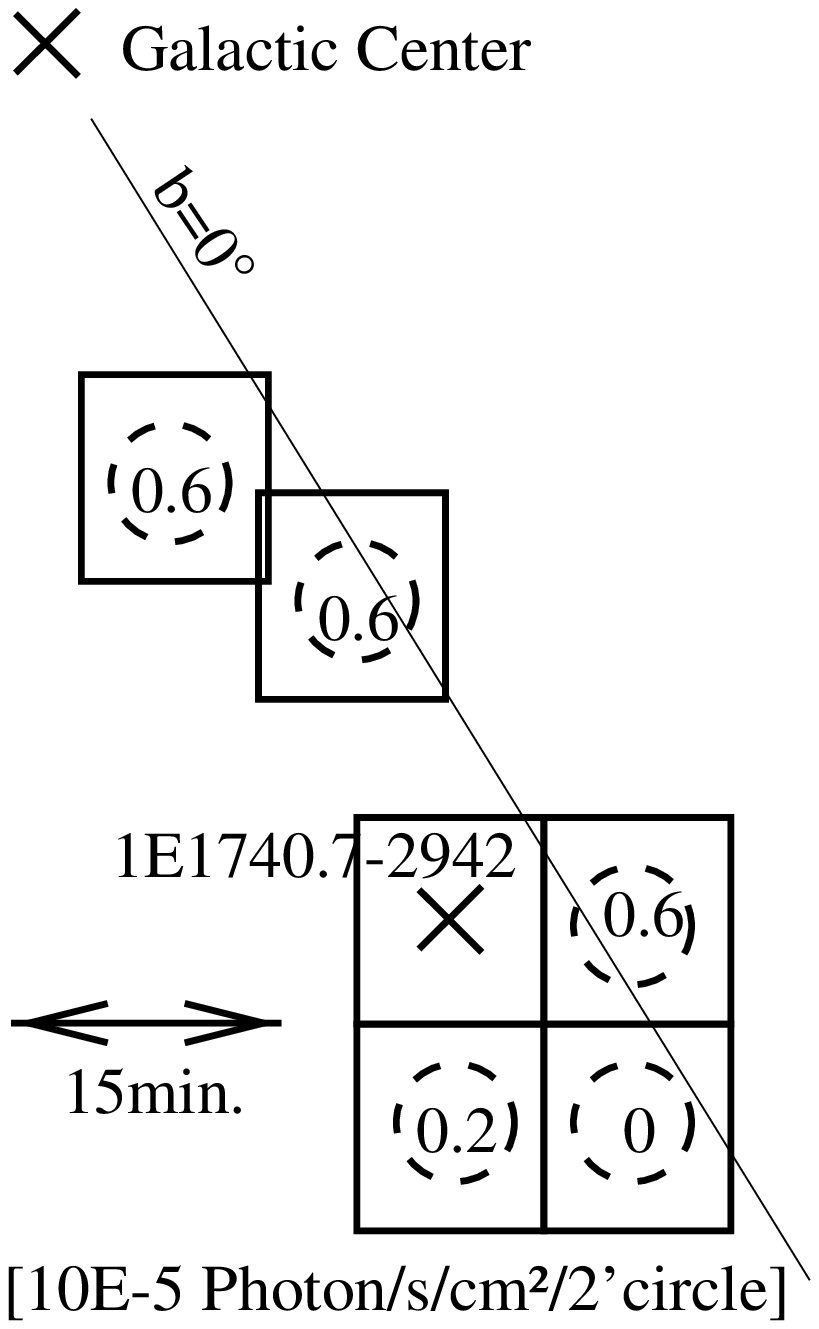,width=0.7\textwidth,clip=}}
  \caption[f5.eps]
{Iron line distribution in the Galactic Center region near {\GA}
estimated by a spectral analysis in a 4$'$ radius circle of each SIS chip
 for the data taken in 1993 (see Table~\ref{tbl:obslog}).
 The best-fit flux of the neutral iron 6.4 keV line
 is normalized to the value in a 2$'$ radius region
 with an assumption of the uniform distribution of diffuse 6.4 keV line
 in the circle.
   These are given in the circles on the figure with the unit of
 10$^{-5}$~photon~s$^{-1}$~cm$^{-2}$/(2$'$-radius circle).\\
\label{fig:iron-diffuse}
}
\end{figure}

\begin{figure}
\centering
  \centerline{\psfig{file=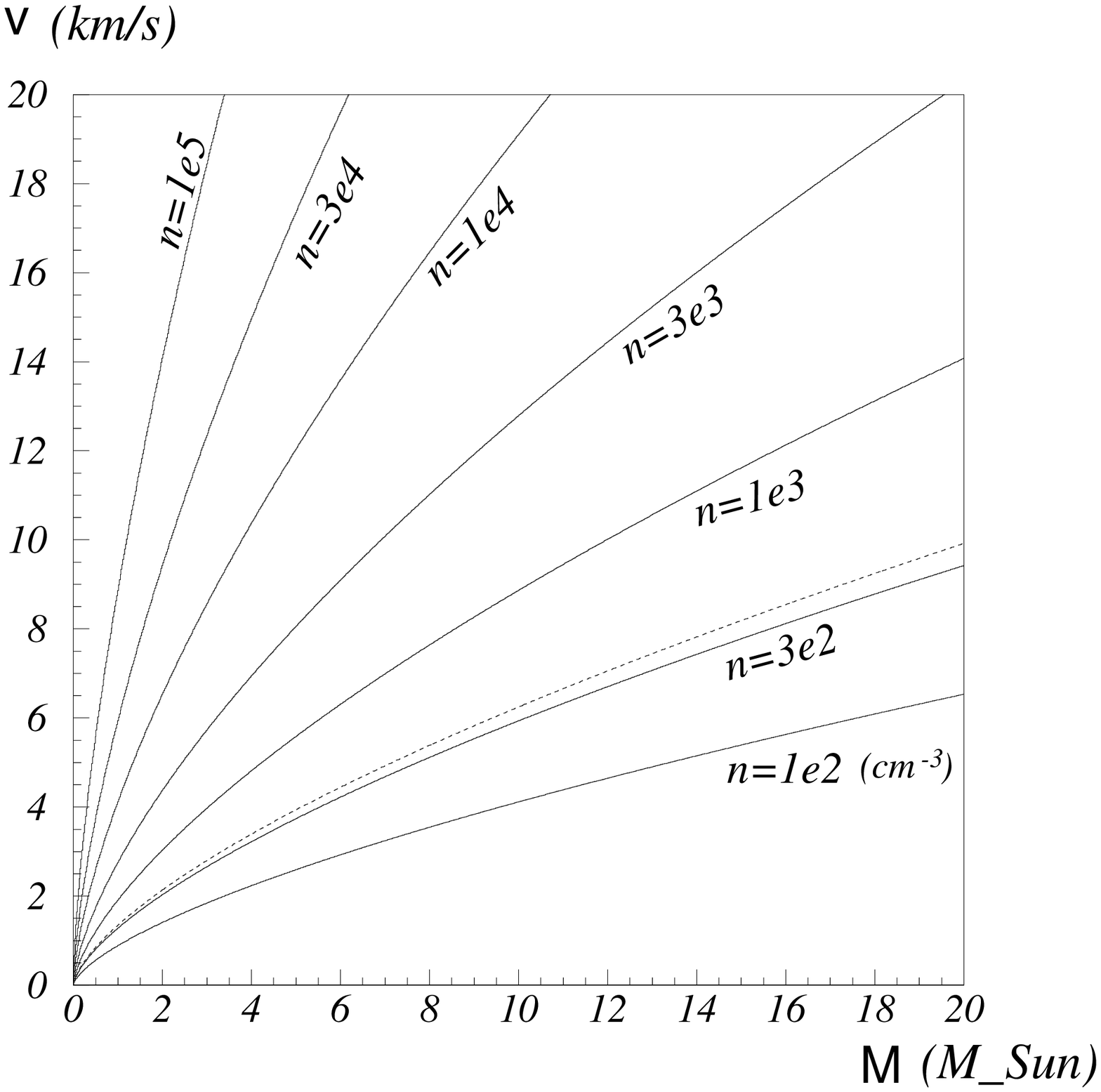,width=\textwidth,clip=}}
  \caption[f6.eps]
{The relation between the mass ($M$) and the velocity ($v$) of the central source
 for certain cloud density ($n$) to explain the observed luminosity of {\GA}
 in the Bondi-Hoyle scenario (solid lines).
  The dashed line shows the case of the estimated upper limit of the density
 (eq. [2] with $Z_{\rm Fe}=2$).
\label{fig:bondi-prm}}
\end{figure}

\small

\begin{table}
\begin{center}
\caption[Observational log]
{Observational log.\label{tbl:obslog}
}

\begin{tabular}{cccclllcc}
\tableline
\tableline
Phase & Date\tablenotemark{a} & R.A.\tablenotemark{b} & DEC.\tablenotemark{b} & Bit-Rate & \multicolumn{2}{c}{Data Mode} & \multicolumn{2}{c}{Exposure\tablenotemark{c}} \\
      &          &        &         &          & GIS& \multicolumn{1}{c}{SIS}  & GIS & SIS \\
\tableline 
PV     & 1993/09/26    & \hms{17}{43}{30} & \dmnos{-29}{50} & High  & PH & 4CCD Faint & 24 & 24 \\
       &               &           &           & Medium& PH & 1CCD Bright& 10 & 12 \\
\tableline 						        
PV\tablenotemark{d} & 1993/10/03 & \hms{17}{45}{30} & \dmnos{-29}{14} & High  & PH & 4CCD Faint &  8 &  6 \\
       &               &           &           & Medium& PH & 4CCD Bright& 15 & 14 \\
\tableline 						        
PV\tablenotemark{d} & 1993/10/04 & \hms{17}{44}{00} & \dmnos{-29}{20} & High  & PH & 4CCD Faint & 18 & 12 \\
       &               &           &           & Medium& PH & 4CCD Bright&  3 &  3 \\
\tableline 						        
AO-2   & 1994/09/08    & \hms{17}{43}{30} & \dmnos{-29}{50} & High  & PH & 2CCD Faint & 12 & 12 \\
       &               &           &           & Medium& PH & 1CCD Faint &  5 &  5 \\
\tableline 						        
AO-2   & 1994/09/12    & \hms{17}{43}{30} & \dmnos{-29}{50} & High  & PH & 2CCD Faint &  9 &  9 \\
       &               &           &           & Medium& PH & 1CCD Faint &  4 &  4 \\
\tableline 
AO-3\tablenotemark{e}  & 1995/03/27    & \hms{17}{44}{10} & \dmnos{-30}{03} & High  & PH & \multicolumn{1}{c}{---} & 24 & -- \\
       &               &           &           & Medium& PH & \multicolumn{1}{c}{---} & 16 & -- \\
\tableline 
AO-5\tablenotemark{e}   & 1997/03/21    & \hms{17}{44}{59} & \dmnos{-29}{46} & High  & PH & \multicolumn{1}{c}{---} & 37 & -- \\
       &               &           &           & Medium& PH & \multicolumn{1}{c}{---} & 49 & -- \\
\tableline 
\end{tabular}
\end{center}

\tablenotetext{a}{year/month/day, in Universal Time.}
\tablenotetext{b}{Coordinate of detector center (J2000).}
\tablenotetext{c}{unit of kilo-second.}
\tablenotetext{d}{These pointings do not include 1E 1740.7$-$2942 in their fields of view.
They are used for background estimation in Sec.~\ref{sec:1740-variability}.}
\tablenotetext{e}{{\GA} is out of the SIS fields of view.}
\end{table}

\clearpage

\clearpage

\begin{table}
    {\scriptsize
\begin{center}
  \caption[Best-fit parameters in the fitting with a power-law function]
{Best-fit parameters in the fitting with a power-law function
\label{tbl:fit-vabs}
}
\begin{tabular}{llcccccccc}
\tableline
\tableline
     &     & \multicolumn{6}{c}{Absorbed Power-law Model} & \multicolumn{2}{c}{6.4~keV-line Inclusive}\\
Phase &  & $\Gamma$\tablenotemark{a} & ${I_{\rm 1keV}}$\tablenotemark{b} & {\NH}\tablenotemark{c} & {\NFe}\tablenotemark{d} & $\chi^2$/d.o.f. & ${F_{\rm X}}$\tablenotemark{e} & ${L_{\rm 6.4keV}}$\tablenotemark{f} & ${EW_{\rm 6.4keV}}$\tablenotemark{g} \\
\tableline
PV   & GIS & $1.28\pm 0.11$         & $4.8^{+1.1}_{-0.9}$ & $10.8\pm 1.1$       & $6.7\pm 2.0$       &  89.2/69 & 2.16& $6\pm 8$  & $17\pm 22$\\
     & SIS & $1.01^{+0.15}_{-0.14}$ & $3.0^{+1.0}_{-0.7}$ & $8.0^{+1.6}_{-1.5}$ & $14.0\pm 2.8$      & 136.4/115& 2.14& $5\pm 7$  & $13\pm 18$\\
 &GIS$+$SIS& $1.18\pm 0.09$         & $4.0^{+0.7}_{-0.6}$ & $9.8\pm 0.9$        & $9.3\pm 1.6$       & 256.0/188& 2.15& $5\pm 5$  & $13\pm 14$\\
\tableline      
AO-2 & GIS & $1.29\pm 0.13$         & $3.3^{+1.0}_{-0.7}$ & $10.6^{+1.4}_{-1.3}$& $7.5\pm 2.3$       &  73.4/63 & 1.46& $<6.3$    & $<25$     \\
     & SIS & $1.04^{+0.14}_{-0.13}$ & $2.3^{+0.7}_{-0.5}$ & $8.0^{+1.5}_{-1.4}$ & $11.7\pm 2.8$      &  75.4/63 & 1.62& $<4.8$    & $<17$     \\
 &GIS$+$SIS& $1.24\pm 0.09$         & $3.2^{+0.6}_{-0.5}$ & $9.8\pm 1.0$        & $9.1^{+1.8}_{-1.7}$& 238.9/130& 1.51& $<4.3$    & $<15$    \\
\tableline      
AO-3 & GIS & $1.23^{+0.29}_{-0.28}$ & $2.0^{+1.4}_{-0.8}$ & $11.2^{+3.0}_{-2.9}$& $4 \pm 5$          &  46.6/44 & 0.96& $<9$      & $<56$     \\
\tableline      
AO-5 & GIS & $0.91\pm 0.10$         & $2.3^{+0.5}_{-0.4}$ & $8.4\pm 1.0$        & $10.4\pm 1.8$      & 131.6/89 & 2.07& $0.7\pm 7$& $2\pm 20$ \\
\tableline

\end{tabular}
\end{center}

\tablenotetext{a}{Photon index.}
\tablenotetext{b}{Intensity of continuum at 1keV [$10^{-2}$photon\ keV$^{-1}$\ cm$^{-2}$\ s$^{-1}$] (absorption corrected).}
\tablenotetext{c}{Hydrogen column density [$10^{22}$ {\NHUNIT}].}
\tablenotetext{d}{Iron column density [$10^{18}$ {\NFeUNIT}].}
\tablenotetext{e}{Observed flux with 2--10 keV [$10^{-10}$ {\FLUXUNIT}].}
\tablenotetext{f}{Luminosity of 6.4keV-line [$10^{-5}$Photon\ cm$^{-2}$\ s$^{-1}$] (absorption corrected).}
\tablenotetext{g}{Equivalent width of 6.4~keV-line [eV].}

    }

\end{table}

\end{document}